\documentclass[amsmath,amssymb,pra,twocolumn]{revtex4}
\usepackage[cp1251]{inputenc}
\usepackage[english]{babel}
\usepackage{color}
\usepackage{graphicx}
\usepackage{dcolumn}
\usepackage{braket}
\usepackage{amsmath}

\usepackage{natbib}

\begin{document}
\bibliographystyle{unsrt}

\title{Schmidt-mode analysis of quadrature entanglement in superpositions of two-mode multiphoton states}

\author{M. V. Fedorov}
\email{fedorovmv@gmail.com}
\address{$^1$A.M.~Prokhorov General Physics Institute,
 Russian Academy of Sciences, 38 Vavilov st., Moscow, 119991, Russia}
 \address{$^2$National Research University Higher School of Economics, 20 Myasnitskaya Ulitsa,
Moscow, 101000, Russia}

%\date{\today}

\begin{abstract}
The  Schmidt-decomposition formalism is proposed to be used for evaluation of the degree of quadrature entanglement in two-mode multiphoton states.
\end{abstract}

\maketitle

\section{Introduction}
As known, states of photons can be characterized in quantum electrodynamics either by their state vectors or by wave functions depending on variables. The simplest example is the state of a single photon with, e.g., horizontal polarization ($H$). Its state vector is $\ket{1_H}\equiv a_H^\dag\ket{0}$ where $a_H^\dag$ is the photon creation operator in the $H$-mode and $\ket{0}$ is the vacuum state. The single-photon wave function of the polarization variable $\sigma$ is given by $\psi(\sigma)=\braket{\sigma|1_H}=\delta_{\sigma,H}$. Two-photon polarization states are characterized by two discrete polarization variables $\sigma_1$ and $\sigma_2$, each of which can take only two ``values", $H$ or $V$ (vertical). Depending on distribution of two photons between two modes $H$ and $V$, biphoton states can be either entangled or not in polarization variables \cite{NJP}. In addition to polarization, photons can have other degrees of freedom, such as frequencies or directions of propagation, and their wave functions can depend on the corresponding variables, discrete or continuous. Entanglement of all such biphoton states can be characterized straightforwardly by the method known as the Schmidt-mode decomposition \cite{Schm,grobe,LE,PRA-08,Straupe,CP,PRA-16}. In the case of multiphoton states, the number of variables grows dramatically as each photon has its own variable for each of its degrees of freedom. The case of multiphoton polarization (two-mode) states was considered recently in the work \cite{Arx}. In spite of serious complications related to a growing number of polarization variables, it appears possible to find the Schmidt entanglement parameters for such states and in this way to evaluate their degree of entanglement with respect to division of pure multiphoton states for  various pairs of subsystems with smaller numbers of photons. As a whole, the Schmidt-mode analysis appears to be a very powerful method of entanglement characterization for both biphoton and multiphoton states with respect to the photon's polarization, frequency and angular variables.

A very different approach to analyzing entanglement of photons is related to modeling their states by the states of equivalent oscillators with wave functions depending on continuous quadrature variables \cite{Heitler} (see definitions in the following section). Such description gives rise to the concept of entanglement in continuous quadrature variables. This is a very popular topic of investigations and there are many works on this subject \cite{Ou,J-Zhang,Bowen,Wenger,Adamyan,Y-Zhang,Manko,Serna}. But it seems that there were no attempts to apply the ideas of the Schmidt-mode decomposition to analysis of quadrature  entanglement in biphoton and multiphoton states. Development of this approach is the main goal of this work. As shown below, the quadrature representation of two-mode states of photons appears to be a very appropriate object for fruitful application of the Schmidt-mode analysis.

\section{Single-mode quadrature variables and wave functions.}
Let $a^\dag$ and $a$ be the quantum-electrodynamic (QED) creation and annihilation operators of a photon in some given mode. Then, the photon quadrature variables for this mode are defined as
\begin{equation}
 \label{def}
 x=\frac{a+a^\dag}{\sqrt{2}}\;{\rm and}\;p=-i\frac{a-a^\dag}{\sqrt{2}}=-i\frac{d}{dx}.
\end{equation}
These definitions provide a transition from the QED description of photon states to their oscillator representation \cite{Heitler}. In this representation the continuous dimensionless variable $x$ is the oscillator coordinate, and its wave functions $\psi_n(x)$ ($n=0,\,1,\,2,\,...$) obey the equation
\begin{equation}
 \label{Schr}
 \frac{1}{2}\left(-\frac{d^2}{dx^2}+x^2\right)\psi_n(x)=\left(n+\frac{1}{2}\right)\psi_n(x),
\end{equation}
where
\begin{equation}
 \label{Herm}
 \psi_n(x)=\frac{1}{\sqrt{2^nn!\sqrt{\pi}}}e^{-x^2/2}H_n(x),
\end{equation}
and $H_n(x)$ are the Hermit polynomials; defined in this way, the oscillator wave functions $\psi_n(x)$  are normalized and orthogonal to each other
\begin{equation}
 \label{norm}
 \int dx \psi_n(x)\psi_{n^\prime}(x)=\delta_{n,n^\prime}.
\end{equation}

The $n^{th}$ levels of the oscillator correspond to the $n$-photon QED states $\ket{n}=\frac{(a^\dag)^n}{\sqrt{n!}}\ket{0}$. In other words, in terms of quadrature variables, the  $n$-photon states are characterized by the oscillator wave function $\psi_n(x)$. The ground state of the oscillator and its wave function $\psi_0(x)$ correspond to the vacuum state in the chosen mode $\ket{0}_{\rm mode}$ (whereas in the QED picture $\ket{0}$ means usually ``no photons in any modes at all").

A general form of the quadrature wave function of an arbitrary state of photons in the given mode has the form
\begin{equation}
 \label{arb-Hsb-quadr}
 \Psi_{\rm Hsb}(x)=\sum_{n=0}^{\infty}A_n\psi_n(x),
\end{equation}
with the expansion coefficients $A_n$ obeying the the normalization condition $\sum_n|A_n|^2=1$.

The form (\ref{arb-Hsb-quadr}) of the quadrature wave function  corresponds to the Heisenberg representation in which wave functions do not depend on time and all the time dependencies are present in operators like that of the electric field strength. The situation is reversed in the Schr${\rm \ddot{o}}$dinger representation where the wave function (\ref{arb-Hsb-quadr}) is replaced by
\begin{equation}
 \label{arb-Sch-quadr}
 \Psi_{\rm Sch}(x;t,{\vec r})=\sum_{n=0}^{\infty}A_n\,e^{in(\omega t-{\vec k}{\vec r})}\,\psi_n(x),
\end{equation}
where $\omega$ and $\vec k$ are the photon's frequency and wave vector.

\section{Two-mode quadrature variables and wave functions.}
Let us consider now the case of two orthogonal modes and two-mode states and wave functions. Specifically, these modes can be characterized by different orthogonal polarizations, e.g., the horizontal ($H$) and vertical ($V$) ones. Alternatively, the modes can correspond to two different propagation directions, or two different frequencies of photons. By returning to two polarization modes, let frequencies $\omega$ and wave vectors $\vec k$ of all photons in these modes be identical. Let the quadrature variables in the $H$-mode be the same as previously,
\begin{equation}
 \label{H-mode}
 x=\frac{a_H+a_H^\dag}{\sqrt{2}}\;{\rm and}\;p=-i\frac{a_H-a_H^\dag}{\sqrt{2}}=-i\frac{\partial}{\partial x},
\end{equation}
and in the $V$-mode be given by
\begin{equation}
 \label{V-mode}
 y=\frac{a_V+a_V^\dag}{\sqrt{2}}\;{\rm and}\; q=-i\frac{a_V-a_V^\dag}{\sqrt{2}}=-i\frac{\partial}{\partial y}.
\end{equation}
With the quadrature wave function in both modes taken in the general form of the type (\ref{arb-Sch-quadr}), the two-mode wave function is given by
\begin{gather}
 \nonumber
 \Psi_{\rm Sch}(x,y;t,{\vec r})=\Psi_{\rm Sch}(x;t,{\vec r})\Psi_{\rm Sch}(y;t,{\vec r})=\\
 \label{HV}
 \sum_{n_H=0}^{\infty}\sum_{n_V=0}^{\infty}A_{n_H}\,B_{n_V}\,e^{i(n_H+n_V)(\omega t-{\vec k}{\vec r})}\,\psi_{n_H}(x)\psi_{n_V}(y)
\end{gather}
with $\sum_{n_H}|A_{n_H}|^2=\sum_{n_V}|B_{n_V}|^2=1$.

A special class of such two-mode quadrature wave functions is that of stationary states. Such states and wave functions arise in the cases of a given finite total number of photons in two modes, $n_H+n_V=N$.  In this case the two-mode wave function (\ref{HV}) takes the form
\begin{gather}
 \label{st-st}
 \Psi_{\rm Sch}^{(N)}(x,y;t,{\vec r})=e^{iN(\omega t-{\vec k}{\vec r})}\Psi_{\rm Hsb}^{(N)}(x,y),
\end{gather}
where
\begin{equation}
 \label{Psi-N-Hsb}
 \Psi_{\rm Hsb}^{(N)}(x,y)=\sum_{n=0}^N C_n\,\psi_{n}(x)\psi_{N-n}(y),
\end{equation}
with
\begin{equation}
 \label{Cn(N)}
 C_n=A_nB_{N-n}
\end{equation}
and the normalization conditions for $A_n$ and $B_n$ replaced by $\sum_{n=0}^N|C_n|^2=1.$

\section{Quadrature entanglement and Schmidt-mode analysis}
In the case of the general-form two-mode wave function (\ref{HV}) its dependence on two variables $x$ and $y$ is factorized and, hence, with respect to quadrature variables this state is disentangled. For entanglement the coefficients $\{A_n\}$ and $\{B_n\}$ in the expansions of Equation (\ref{HV}) have to be correlated with each other in some way. One way of correlations between these coefficients occurs in the case of stationary states (\ref{Psi-N-Hsb}) and is given by Equation (\ref{Cn(N)}).

As for the degree of entanglement, as known, for states with wave functions depending on two variables their entanglement can be straightforwardly evaluated in terms of the Schmidt modes and Schmidt decompositions. It's known also that the Schmidt modes can be found from integral equations, kernel of which is given by the wave function of two arguments itself. For an arbitrary wave function $\Psi(x,y)$ equations for adjoint Schmidt modes $\phi_n(x)$ and $\chi_n(y)$ are given by
\begin{equation}
 \label{Schm-m}
  \begin{matrix}
  \int dy \Psi(x,y)\chi_n^*(y)=\sqrt{\lambda_n}\phi_n(x),\\
  \,\\
  \int dx \Psi(x,y)\phi_n^*(x)=\sqrt{\lambda_n}\chi_n(y)
  \end{matrix}
\end{equation}

For the quadrature wave function of the stationary state (\ref{Psi-N-Hsb}), because of orthogonality of the oscillator eigenfunctions (\ref{norm}), the adjoint Schmidt modes are $\phi_n=e^{\varphi_n/2}\psi_n(x)$ and $\chi_n=e^{\varphi_n/2}\psi_{N-n}(y)$, where $\varphi_n$ is the phase of $C_n$. In this case the expression (\ref{Psi-N-Hsb}) for the wave function $\Psi^{N}_{\rm Hsb}(x,y)$  represents itself the Schmidt decomposition of this wave function with the decomposition coefficients $\sqrt{\lambda_n}=|C_n|$.

As known, in terms of the Schmidt decomposition and its parameters, the degree of entanglement of bipartite states can be characterized by the Schmidt entanglement parameter $K$ defined as the inverse sum of squared parameters $\lambda_n$. The case $K=1$ corresponds to disentagled states. In other cases the difference $K-1$ can be considered as the measure of entanglement. For the stationary two-mode quadrature states (\ref{Psi-N-Hsb}) this definition yields
\begin{equation}
 \label{K}
 K=\frac{1}{\sum_n\lambda_n^2}=\frac{1}{\sum_{n=0}^N|C_n|^4}.
\end{equation}
With normalization of the constants $C_n$ taken into account, it's clear from Equation (\ref{K}) that the entanglement parameter $K$ is maximal if all constants $C_n$ are equal to each other and equal to $C_n=1/\sqrt{N+1}$ which gives $K_{\max}=N+1$. Thus, the maximal degree of the quadrature entanglement  of the state (\ref{Psi-N-Hsb}) equals $K_{\max}-1=N$, and grows linearly with gowing number of terms in the superposition (\ref{Psi-N-Hsb}). For comparison, as found in the work \cite{Arx}, the maximal achievable degree of polarization entanglement in the state
$\ket{n_H,(N-n)_V}$
equals approximately $\sqrt{N}$, i.e. it grows much slower with growing $N$ than the degree of quadrature entanglement in the state (\ref{Psi-N-Hsb}).

Another special class of states with equally high quadrature entanglement is that with the non-stationary two-mode quadrature wave function of the form
\begin{equation}
 \label{2n-non-st}
 \Psi(x,y)=\sum_n C_n e^{2in(\omega t-{\vec k}{\vec r})}\psi_n(x)\psi_n(y)
\end{equation}
with $C_n=A_nB_n$ and $\sum_n|C_n|^2=1$.
This sum can be considered again as the Schmidt decomposition with the pairs of Schmidt modes $\{e^{in(\omega t-{\vec k}{\vec r})+i\varphi_n/2}\psi_n(x),\, e^{in(\omega t-{\vec k}{\vec r})+i\varphi_n/2}\psi_n(x)\}$, with $\varphi_n$ representing phases of the coefficients $C_n$, and with the Schmidt entanglement parameter $K$ given by
\begin{equation}
 \label{K-2n-non-st}
 K=\frac{1}{\sum_n|C_n|^4}.
\end{equation}
Evidently, with sufficiently large number of non-zero terms in the wave function (\ref{2n-non-st}) and, consequently, sufficiently small values of $|C_n|^2$, the degree of quadrature entanglement determined by the parameter $K$ of Equation (\ref{K-2n-non-st}) can be very high.

Note also that in the QED notations the wave function (\ref{2n-non-st}) corresponds to the state vector
\begin{equation}
 \label{st-vect-mult}
 \ket{\Psi}=\sum_n C_n \frac{(a_H^\dag)^n(a_V^\dag)^n}{n!}\ket{0}
\end{equation}
In terms of wave functions with polarization variables each $n^{th}$ term in this sum corresponds to a function of $2n$ polarization variables $\sigma_i^{H}$ and $\sigma_i^{V}$ with $1\leq i\leq n$ (one variable per one photon). The sum of wave functions with different numbers of variables makes impossible application to such states of the Schmidt-decomposition method. In contrast, the quadrature wave function of the same state (\ref{2n-non-st}) remains the function of only two arguments, $x$ and $y$, at any number of terms in the sum over $n$, which makes the Schmidt-mode analysis perfectly applicable. In other words in the cases of two-mode multiphoton states their quadrature description represents a much more appropriate object for their Schmidt-mode characterization than the polarization-variable representation.

\section{Examples}

1. {\bf Quadrature entanglement of single-photon and vacuum two-mode states}.
\begin{equation}
\label{01-10}
 \ket{\Psi}=\cos\alpha\ket{0_H,1_V}+\sin\alpha\ket{1_H,0_V}\equiv a_\alpha^\dag\ket{0},
\end{equation}
where $a_\alpha^\dag$ is the QED creation operator for a photon with the polarization turned for the angle $\alpha$ with respect to the horizontal direction. The quadrature-variable wave function of the state (\ref{01-10}) is given by
\begin{gather}
 \nonumber
 \Psi(x,y)=\cos\alpha\,\psi_0(x)\psi_1(y)+\sin\alpha\,\psi_1(x)\psi_0(y)=\\
 \label{WF-01-10}
 =\frac{1}{\sqrt{\pi}}e^{-(x^2+y^2)/2}(\cos\alpha\; x+\sin\alpha\; y).
\end{gather}
The degree of quadrature entanglement is
\begin{equation}
 \label{K-01-HV}
 K^{(01)}(\alpha)=\frac{1}{\sin^4\alpha+\cos^4\alpha}.
\end{equation}
The maximal value of the entanglement parameter $K^{(01)}(\alpha)$ is achieved at $\alpha=\pi/4$ and $K^{(01)}_{\max} =2$. At $\alpha=0$ or $\pi/2$ the state (\ref{01-10}) is disentangled, $K^{(01)}(0)=K^{(01)}(\pi/2)=1$.

Note, that the specific positions of maximal and missing quadrature entanglement depend on the choice of two orthogonal modes in which the two-mode wave function is defined or measurements are assumed to be done. For example, if these modes would correspond to polarizations $\beta$ and $\beta+90^\circ$, the function $K^{(01)}(\alpha)$ would have maximum and minima, correspondingly, at $\alpha=\beta+\pi/4$ and at $\alpha=\beta$ or $\alpha=\beta+\pi/2$. In other words, the quadrature entanglement is not invariant with respect to rotations in the polarization plane perpendicular to the photon wave vectors. Note also, that in contrast to the quadrature entanglement, entanglement in polarization variables is not defined at all in the case of one-photon states because such states are characterized by only one polarization variable of the only photon, and there is no partner-variable to be entangled with.

2. {\bf Quadrature entanglement of two-mode biphoton states (qutrits)}.
\begin{gather}
\nonumber
 \ket{\Psi}=C_1\ket{2_H,0_V}+C_2\ket{1_H,1_V}+C_3\ket{0_H,2_V}\equiv\\
 \label{biph}
 \left(C_1\frac{(a_H^\dag)^2}{\sqrt{2}}+C_2a_H^\dag a_V^\dag+C_3\frac{(a_V^\dag)^2}{\sqrt{2}}\right)\ket{0},
\end{gather}
Such states are known as qutrits and they are states of the type (\ref{Psi-N-Hsb}). The degree of their quadrature entanglement is characterized by the entanglement parameter $K$ of Equation (\ref{K})
\begin{equation}
 \label{K-qtr}
 K^{\rm qtr}_{quadr}=\frac{1}{|C_1|^4+|C_2|^4+|C_2|^4}.
\end{equation}
Entanglement is maximal at $|C_1|=|C_2|=|C_3|=1/\sqrt{3}$ and $K^{\rm qtr}_{\max}=3$. For comparison, for the same states but considered in polarization rather than quadrature variables, the Schmidt entanglement parameter is given by \cite{NJP,JETP}
\begin{equation}
 \label{K-pol-qtr}
 K^{\rm qtr}_{pol} =\frac{2}{2-|2C_1C_3-C_2^2|^2}.
\end{equation}
The maximal value of $K^{\rm qtr}_{pol}$ is $K^{\rm qtr}_{pol\,\max}=2$ and at $|C_1|=|C_2|=|C_3|=1/\sqrt{3}$, as follows from (\ref{K-pol-qtr}), $K^{\rm qtr}_{pol}=\frac{18}{17}$, which corresponds to the degree of entanglement $^{\rm qtr}_{pol}-1=\frac{1}{18}\ll 1$, whereas in the same case $K^{\rm qtr}_{quadr}=3$.

Another example interesting for comparison of the polarization and quadrature entanglement is that of the state $\ket{1_H,1_V}$. Owing to  obligatory symmetry of the corresponding polarization wave function, this state is maximally entangled, with the Schmidt parameter equal to $K_{pol}^{\ket{1H,1V}}=2$. In the same time the quadrature wave function of the state $\ket{1_H,1_V}$ is given by a simple product of two identical single-photon wave functions $\Psi_{quadr}^{\ket{1H,1V}}(x,y)=\psi_1(x)\psi_1(y)$. Clearly enough, this is the wave function of a disentangled state with  $K_{quadr}^{\ket{1H,1V}}=1$.

3. {\bf Quadrature entanglement of the two-mode squeezed-vacuum state}

As known \cite{Caves,MW}, the two-mode squeezed-vacuum state is defined as
\begin{equation}
 \label{sq-vac-st}
 \ket{\Psi_{sq. vac}} = \exp\left [ r\left(a_H a_V e^{-2i\varphi}+a_H^\dag a_V^\dag e^{2i\varphi}\right)\right ]\ket{0}.
\end{equation}
Decomposition of this state in a series of two-mode Fock sates $\ket{n_H,n_V}$ is known also. As it was shown in the work \cite{Caves} (equations (3.66) and (4.39)), decomposition of the state (\ref{sq-vac-st}) in a series of two-mode states contains only Fock states with equal numbers of photons in two modes, $n_H=n_V\equiv n$ and, specifically, it has the form 
\begin{equation}
 \label{sq-vac-st}
 \ket{\Psi_{\rm sq. vac}} = (\cosh{r})^{-1}\sum_{n=0}^\infty\left(-e^{2i\varphi}\,\tanh{r}\right)^n\ket{n,n},
\end{equation}
where $|\tanh{r}|<1$. Evidently, the state $\ket{\Psi_{\rm sq. vac}}$ belongs to the class of states (\ref{2n-non-st}) with
\begin{equation}
 \label{Schm-parm}
 |C_n^{\,\rm sq-vac}|=\sqrt{\lambda_n}=\frac{(\tanh{r})^n}{\cosh{r}}.
\end{equation}
The two-mode quadrature wave function of the state $\ket{\Psi_{\rm sq. vac}}$ is
\begin{equation}
 \label{wf-sq-vac}
 \Psi^{\,\rm sq-vac}_{quadr}(x,y)= e^{in(2\varphi+\pi)}\sqrt{\lambda_n}\,\psi_n(x)\psi_n(y),
\end{equation}
and it represents the Schmidt decomposition of the quadrature two-mode squeezed-vacuum wave function.

As it has to be, the state $\ket{\Psi_{\rm sq. vac}}$ obeys the unit-normalization condition
\begin{gather}
 \nonumber
 \braket{\Psi_{\rm sq. vac}|\Psi_{\rm sq. vac}}=\sum_{n=0}^\infty|C_n^{\,\rm sq-vac}|^2=\sum_{n=0}^\infty\lambda_n=\\
 \label{norm}
 =\frac{\sum_{n=0}^\infty(\tanh r)^{2n}}{\cosh^2 r}=\frac{1}{\cosh^2 r(1-\tanh^2 r)}=1.
\end{gather}
Note that the sum over $n$ in the second line of this equation is the geometric progression with the common ratio $(\tanh{r})^2<1$. In the following calculation of the Schmidt entanglement parameter $K$ we will meet a similar progression but with the common ratio $(\tanh{r})^4<1$.

By general rules, the Schmidt entanglement parameter of the two-mode quadrature state $\ket{\Psi_{\rm sq. vac}}$ is given by
\begin{gather}
 \nonumber
 K^{\,\rm sq-vac}_{quadr}=\frac{1}{\sum_{n=0}^\infty|C_n^{\,\rm sq-vac}|^4}=\frac{1}{\sum_n\lambda_n^2}=\\
 \label{K-sq-vac}
\frac{\cosh^4 r}{\sum_{n=0}^\infty (\tanh r)^{4n}}=\cosh^4 r (1-\tanh^4 r)=\cosh 2r.
\end{gather}
Thus, the degree of quadrature entanglement of the two-mode squeezed-vacuum state is strongly dependent on the degree of  squeezing, and for a highly squeezed state the entanglement Schmidt parameter grows exponentially with  a growing extent of squeezing, $K^{\,\rm sq-vac}_{quadr}\approx \frac{1}{2}e^{2r}$ at $r\gg 1$.

Finally, it may be interesting to compare the derived result with the total number $N$ of photons in the state $\ket{\Psi_{\rm sq. vac}}$ given by
\begin{gather}
 \nonumber
 N_{\rm tot}^{\,\rm sq-vac}=\frac{2}{\cosh^2 r}\sum_{n=0}^\infty n (\tanh r)^{2n}=\frac{2\tanh^2r}{\cosh^2r(1-\tanh^2r)^2}\\
 \label{N}
 =2\sinh^2r=K^{\,\rm sq-vac}_{quadr}-1.
\end{gather}
Though the final expressions of Equations (\ref{K-sq-vac}) and (\ref{N}) are not exactly identical, in the case of strong squeezing, $r\gg 1$, they coincide and they are both exponentially high
\begin{equation}
 \left.N_{\rm tot}^{\,\rm sq-vac}\right|_{r\gg 1}=\left.K^{\,\rm sq-vac}_{quadr}\right|_{r\gg 1} =\frac{e^{2r}}{2}.
 \label{asympt}
\end{equation}

\section{Conclusion}

The presented general analysis and given examples indicate fruitfulness of application of the Schmidt-mode formalism for characterization of quadrature entanglement in both a few- and multi-photon two-mode states. As shown, in comparable cases there is no direct parallelism between the quadrature entanglement and entanglement in polarization variables. In some cases the quadrature entanglement can be rather high whereas entanglement in polarization variables very low, and in some other cases the situation is reversed.

A remarkable feature of two-mode quadrature wave function is that for them the Schmidt-mode analysis is equally well applicable both to states with given numbers of photons and to superpositions of states with different numbers of photons. This feature differs significantly quadrature two-mode wave functions from similar polarization wave functions. The key point for this difference is in numbers of variables. Polarization wave functions of states with different numbers of photons depend on different numbers of variables, which excludes the use of the Schmidt mode analysis for superpositions of such wave functions and such states. In contrast to this, the two-mode quadrature wave functions of any multiphoton states depend only on two variables $x$ and $y$, which makes arbitrary superpositions of such wave functions fully appropriate for the Schmidt-mode analysis. As shown, in the cases of superpositions of many two-mode multiphoton states, their  quadrature entanglement can be very high, and the degree of entanglement grows linearly with increasing numbers of terms in these superpositions. Extremely high level of entanglement is shown to occur in the case of the two-mode squeezed-vacuum state, for which the Schmidt entanglement parameter is shown to grow exponentially with a growing degree of squeezing. Applications and aspects of experimental observations will be discussed elsewhere.

\section*{Acknowledgement}
The work is supported by the Russian Foundation for Basic Research, grant 18-02-00634.

\bibliography{text}

\begin{thebibliography}{10}

\bibitem{NJP}
M.~V. Fedorov, P.~A. Volkov, J.~M. Mikhailova, S.~S. Straupe, and S.~P. Kulik.
\newblock Entanglement of biphoton states: qutrits and ququarts.
\newblock {\em New Journ. of Phys.}, 13:83004, 2011.

\bibitem{Schm}
E.~Scmidt.
\newblock Zur theorie der linearen und nichtlinearen lntegralgleichungen.
\newblock {\em Math Ann.}, 63:433, 1951.

\bibitem{grobe}
R.~Grobe, K.~Rzazewski, and J.~H. Eberly.
\newblock Measure of electron-electron correlation in atomic physics.
\newblock {\em J. Phys. B: At. Mol. Opt. Phys.}, 27:L503, 1994.

\bibitem{LE}
C.~K. Law and J.~H. Eberly.
\newblock Analysis and interpretation of high transverse entanglement in
  optical parametric down conversion.
\newblock {\em Phys. Rev. Lett.}, 92:127903, 2004.

\bibitem{PRA-08}
Yu.~M. Mikhailova, P.~A. Volkov, and M.~V. Fedorov.
\newblock Biphoton wave packets in parametric down-conversion: Spectral and
  temporal structure and degree of entanglement.
\newblock {\em Phys. Rev. A}, 78:062327, 2008.

\bibitem{Straupe}
S.~S. Straupe, D.~P. Ivanov, A.~A. Kalinkinand I.~B. Bobrov, and S.~P. Kulik.
\newblock Angular schmidt modes in spontaneous parametric down-conversion.
\newblock {\em Phys. Rev. A}, 83:060302(R), 2011.

\bibitem{CP}
M.~V. Fedorov and N.~I. Miklin.
\newblock Angular schmidt modes in spontaneous parametric down-conversion.
\newblock {\em Contemporary Physics}, 55:94, 2014.

\bibitem{PRA-16}
M.~V. Fedorov.
\newblock Azimuthal entanglement and multichannel schmidt-type decomposition of
  noncollinear biphotons.
\newblock {\em Phys. Rev. A}, 93:033830, 2016.

\bibitem{Arx}
D.~A.~Grigoriev S.~V.~Vintskevich and M.~V. Fedorov.
\newblock Azimuthal entanglement and multichannel schmidt-type decomposition of
  noncollinear biphotons.
\newblock ArXiv:1905.03459v1 [quant-ph], 2019.

\bibitem{Heitler}
W.~Heitler.
\newblock {\em The Quantum Theory of radiation}.
\newblock Oxford, Clarendon Press, 1954.

\bibitem{Ou}
Z.~Y. Ou, S.~F. Pereira, H.~J. Kimble, and K.~C. Peng.
\newblock Realization of the einstein-podolsky-rosen paradox for continuous
  variables.
\newblock {\em Phys. Rev. Lett.}, 68:3663, 1992.

\bibitem{J-Zhang}
Jing Zhang, Changde Xie, and Kunchi Peng.
\newblock Quantum entanglement and squeezing of the quadrature difference of
  bright light fields.
\newblock {\em Phys. Rev. A}, 66:042319, 2002.

\bibitem{Bowen}
W.~P. Bowen, R.~Schnabel, P.~K. Lam, and T.~C. Ralph.
\newblock Experimental investigation of criteria for continuous variable
  entanglements.
\newblock {\em Phys. Rev. Lett}, 90:043601, 2003.

\bibitem{Wenger}
J.~Wenger, A.~Ourjoumtsev, R.~Tualle-Brouri, and P.~Grangier.
\newblock Time-resolved homodyne characterization of individual
  quadrature-entangled pulses.
\newblock {\em Eur. Phys. J. D}, 32:391, 2005.

\bibitem{Adamyan}
H.~H. Adamyan, N.~H. Adamyan, S.~B. Manvelyan, and G.~Yu. Kryuchkyan.
\newblock Quadrature entanglement and photon-number correlations accompanied by
  phase-locking.
\newblock {\em Phys. Rev. A}, 73:033810, 2006.

\bibitem{Y-Zhang}
Y.~Zhang, T.~Furuta, R.~Okubo, K.~Takahashi, and T.~Hirano.
\newblock Experimental generation of broadband quadrature entanglement using
  laser pulses.
\newblock {\em Phys. Rev. A}, 76:012314, 2007.

\bibitem{Manko}
S~N Filippov and V~I Man'ko.
\newblock Optical tomography of fock state superpositions.
\newblock {\em Phys. Scr.}, 76:058101, 2007.

\bibitem{Serna}
F.~A. Dom\'{\i}nguez-Serna, F.~J. Mendieta-Jimenez, and F.~Rojas.
\newblock Relationship between the field local quadrature and the quantum
  discord of a photon-added correlated channel under the influence of
  scattering and phase fluctuation noise.
\newblock {\em Quantum Information Process}, 16:255, 2017.

\bibitem{JETP}
M.~V. Fedorov, P.~A. Volkov, and Yu.~M. Mikhailova.
\newblock Qutrits and ququarts in spontaneous parametric down-conversion,
  correlations and entanglement.
\newblock {\em JETP}, 115:15, 2012.

\bibitem{Caves}
B.L. Schumaker and C.M. Caves.
\newblock New formalism for two-photon quantum optics.
\newblock {\em Phys. Rev. A}, 31:3093, 1985.

\bibitem{MW}
L.~Mandel and E.~Wolf.
\newblock {\em Optical Coherence and Quantum Optics}.
\newblock Cambridge University Press, 1995.

\end{thebibliography}

\end{document}